\documentstyle[epsf]{article}
\begin{document}
\large

\baselineskip 10mm
\begin{center}
{\huge Quantum kinetic theory of shift current \\
electron pumping in semiconductors}
\end{center}

\baselineskip 6mm

\vspace{1mm}
\begin{center}
{\Large Petr Kr\'al$^{\dagger}$}
\end{center}

\begin{center}
{\it Department of Physics, University of Toronto,\\
60 St. George Street, Ontario, Toronto M5S 1A7, Canada}\\
\end{center}

\begin{center}
{\it $^{\dagger}$ Present address:
Department of Chemical Physics,\\
Weizmann Institute of Science,
76100 Rehovot, Israel}
\end{center}

\begin{abstract}
We develop a theory of laser beam generation of {\it shift currents}
in non-centrosymmetric semiconductors. The currents originate when 
the excited electrons transfer between different bands or scatter inside
these bands, and asymmetrically shift their centers of mass in elementary 
cells. Quantum kinetic equations for hot-carrier distributions and 
expressions for the induced currents are derived by nonequilibrium Green 
functions. In applications, we simplify the approach to the Boltzmann 
limit and use it to model laser-excited GaAs in the presence of LO phonon 
scattering. The shift currents are calculated in a steady-state regime.
\end{abstract}

\baselineskip 6mm
%
%  Uncomment out if preprint format required
%
%\pacs{00.00, 20.00, 42.10}

\newpage
 
%%%%%%%%%%%%%%%%%%%%%%%%%%%%%%%%%%%%%%%%%%%%%%%%%%%%%%%%%%%%%%%%%%%%%%%%%%%%
\section{Introduction}
%%%%%%%%%%%%%%%%%%%%%%%%%%%%%%%%%%%%%%%%%%%%%%%%%%%%%%%%%%%%%%%%%%%%%%%%%%%% 
Photovoltaic phenomena in semiconductors can originate in the built-in or 
induced asymmetry or inhomogeneity of their crystal structures \cite{Sturman}. 
In non-centrosymmetric (NCS) crystals, different generation rates for 
carriers at momenta $\pm {\bf k}$ can be induced by asymmetric electron-hole 
scattering and other processes. The resulting momentum imbalance generates 
the so called {\it ballistic current}. Recently \cite{Kurizki,Dupont}, 
this momentum asymmetry of carrier generation in semiconductors was achieved 
by mixing one- and two-photon transitions at frequencies $2\omega_0$ and 
$\omega_0$, respectively. In Ref.\cite{KralCOH}, we describe the effect 
in GaAs in the presence of scattering on LO phonons. We have also suggested 
that the current induced by this two-beam coherent control could drive 
intercalated atoms in carbon nanotubes \cite{PUMA}.

\begin{figure}[htbp]
\vspace*{-26mm}
\hspace*{-10mm}
\epsfxsize=143mm
\vspace*{-30mm}
\epsffile{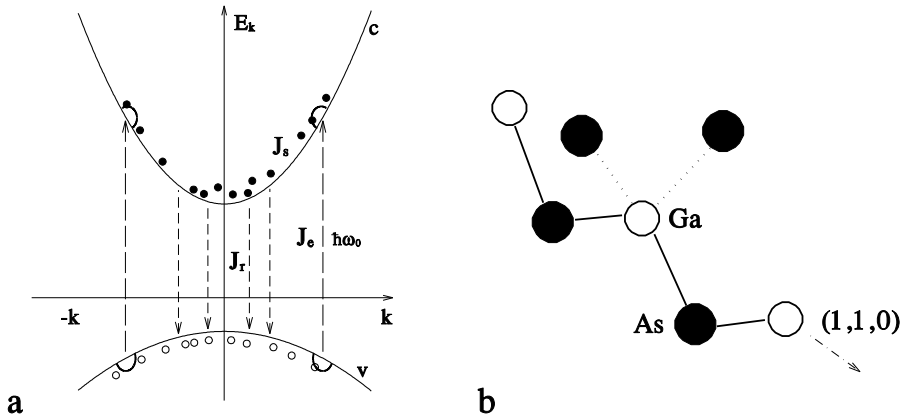}
\vspace*{4mm}
\caption{Scheme of the shift current generation in a NCS semiconductor
is shown in the inset a).
The current has three components, related with the process of electron 
excitation ${\bf J_e}$, scattering ${\bf J_s}$ and recombination 
${\bf J_r}$. They result as a consequence of real space shifts of electrons 
in the elementary cells that undergo the corresponding transitions 
in reciprocal space. In the inset b) we show a small section of a 
zinc-blend structure in GaAs. The excitation conditions and the
generated shift currents ${\bf J_{e,s,r}}$ are described in the text.}
\label{Kral1}
\end{figure}

In bulk NCS semiconductors, light induced interband transitions of electrons 
in reciprocal space are accompanied by their {\it asymmetrical shifts} in the 
real space between atoms in elementary cells.  Similar shifts occur if the 
transitions are induced by scattering or in recombination. First realistic 
description of analogous effects in magnetic materials was given by 
Luttinger \cite{Luttinger}.  The carrier shifts generate the so called 
{\it shift current} \cite{Kraut,Kristoffel,Belinicher,Kurt}, which has 
in general three components, excitation ${\bf J_e}$, scattering ${\bf J_s}$ 
and recombination ${\bf J_r}$, named after the corresponding  processes.
The carrier transitions and the related currents are symbolically sketched in 
Fig.\ref{Kral1}a.  The ballistic current and the excitation part of the shift 
current ${\bf J_e}$ in GaAs were investigated analytically in 
Ref.\cite{Belinicher2}; ${\bf J_e}$ induced by transitions between light 
and heavy hole bands was also studied \cite{Rasulov}. A summary of older 
results about the shift current is presented in Ref.\cite{Sturman}, but 
the derivations of the relevant formulas is less clearly documented.

Here, we develop a quantum kinetic theory for the shift current induced by
laser excitation and scattering. The expressions for the current components 
and the transport equations for carrier populations are derived by  
nonequilibrium Green functions \cite{Rostock} (NGF). In applications, 
we simplify the 
approach to the Boltzmann limit and use it to study optically excited 
GaAs in the presence of scattering by LO phonons.  In our modeling, we 
consider steady-state excitations by a linearly polarized light, and find 
that ${\bf J_e}$ is reasonably large, while ${\bf J_s}$ is two 
orders smaller and ${\bf J_r}$ is negligible. Therefore, {\it continuous 
electron pumping} through the crystal can be achieved. The ultrafast response 
of ${\bf J_e}$, without additional saturation and relaxation tails from
${\bf J_s}$ and ${\bf J_r}$, could be useful in opto-electrical applications.  

The paper is organized as follows. In Sec.2 we describe our model system.
Section 3 is devoted to derivation of expressions for the current components. 
In Sec.4 these expressions are further approximated. Numerical results 
for the hot-electron populations and the induced currents in 
GaAs are presented in Sec.5.

%%%%%%%%%%%%%%%%%%%%%%%%%%%%%%%%%%%%%%%%%%%%%%%%%%%%%%%%%%%%%%%%%%%%%%%%%%%%%%
\section{The system studied }
%%%%%%%%%%%%%%%%%%%%%%%%%%%%%%%%%%%%%%%%%%%%%%%%%%%%%%%%%%%%%%%%%%%%%%%%%%%%%%
The physics of the phenomenon can be understood from Fig.\ref{Kral1}b,
where a segment of the zinc-blend structure for GaAs is shown. The valence 
band states are predominantly localized around the As atoms, while the 
conduction band states are shifted toward the Ga atoms. Therefore, if 
the light is polarized along the (1,1,0) direction, the excited electrons 
transfer from the As atoms at the {\it bottom} to the Ga atoms in the 
middle, giving the excitation current ${\bf J_e}$ in the (0,0,-1) direction 
(negative charge). If the light is polarized in the (1,-1,0) direction, 
electron transitions along bonds orthogonal to the light polarization, 
{\it i.e.} from the As atoms at the {\it top} to the Ga atoms in the 
middle, generate 
${\bf J_e}$ in the (0,0,1) direction.  During relaxation, the excited 
electrons (holes) slightly move their centers of charge and stay close 
to the Ga (As) atoms. Therefore, ${\bf J_s}$ is rather small, as we 
show in our calculations.  On the other hand, scattering redistributes 
the carrier momenta, so that electrons at Ga atoms recombine symmetrically 
with holes at all neighbor As atoms, which gives negligible ${\bf J_r}$. 

%---------------------------------------------------------------------------
\subsection{The model Hamiltonian}
%---------------------------------------------------------------------------
The space shifts of carriers and the related currents can be 
calculated from a combination of interband and intraband transitions. 
The length gauge with the elements of the position operator ${\bf x}$ 
evaluated as in Blount's work \cite{Blount,Ghahramani} gives a convenient
basis for the description \cite{Sipe}. Therefore, we model the 
photo-excited bulk GaAs by the following Hamiltonian \cite{Sipe} 
\begin{eqnarray}
H & = & \sum_{\alpha;{\bf k}} \epsilon_{\alpha}({\bf k})\
        a_{\alpha,{\bf k}}^{\dagger}\ a_{\alpha,{\bf k}}\
 + \sum_{{\bf q}}\ \hbar \omega_{\bf q}\ b^{\dagger}_{\bf q}\ b_{\bf q}
\nonumber \\
 & - &  i~e~{\bf E}(t) \cdot \biggl\{\ \frac12\ \sum_{\alpha;{\bf k}}\ \biggl(
  a_{\alpha,{\bf k}}^{\dagger}\ \frac{\partial a_{\alpha,{\bf k}}}
{\partial {\bf k}} - \frac{\partial a_{\alpha,{\bf k}}^{\dagger}}
{\partial {\bf k}}\ a_{\alpha,{\bf k}} \biggr) -i \sum_{\alpha,\beta;{\bf k}}
{\bf \xi}_{\alpha\beta}({\bf k})\ a_{\alpha,{\bf k}}^{\dagger}\
a_{\beta,{\bf k}} \biggr\}
\nonumber \\
& + & \sum_{\alpha,\beta;{\bf k},{\bf q}}\
M_{\alpha\beta}({\bf k},{\bf k}-{\bf q})\ a_{\alpha,{\bf k}}^{\dagger}
\ a_{\beta,{\bf k}-{\bf q}}\ (b_{\bf q}+b^{\dagger}_{-{\bf q}})\ ,
\label{h}
\end{eqnarray}
where coupling to LO phonons is added. Here the creation (annihilation) 
operators $a_{\alpha , {\bf k}}^{\dagger}$ ($a_{\alpha, {\bf k}}$) 
describe electrons with the band index $\alpha$ at wave vector ${\bf k}$ 
in the Brillouin zone.  The operator $b^{\dagger}_{\bf q}$ ($b_{{\bf q}}$) 
creates (annihilates) phonons with the wave vector ${\bf q}$. The electric 
field is ${\bf E}(t) ={\bf E}_{+\omega_0}(t)~e^{-i\omega_0t} + 
{\bf E}_{-\omega_0}(t) ~e^{+i\omega_0t}$, where ${\bf E}_{+\omega_0}(t)
={\bf E}^{*}_{-\omega_0}(t)$ are complex envelope functions. Spins are
included in the current by a factor of $2$ (see Eqn.(\ref{Jsh})).
 
%---------------------------------------------------------------------------
\subsection{The matrix elements}
%---------------------------------------------------------------------------
The matrix elements ${\bf x}_{\alpha\beta} ({\bf k}, {\bf k}^{'})$ for the 
position operator are defined as \cite{Blount} 
\begin{equation}
\label{rdef}
{\bf x}_{\alpha\beta} ({\bf k}, {\bf k}^{'})
= i~\delta_{\alpha \beta}~\frac{\partial}{\partial {\bf k}}\
  \delta ({\bf k} - {\bf k}^{'})
+ \delta({\bf k} - {\bf k}^{'}) \, \xi_{\alpha \beta}({\bf k})\ ,
\label{xdefi}
\end{equation}
where the functions
\begin{equation}
\xi_{\alpha\beta}({\bf k}) = \xi_{\beta\alpha}^{*}({\bf k}) =
\int\limits_{\mbox{\tiny U.C.}} d{\bf x} \ 
u_{\alpha {\bf k}}^{*}({\bf x}) \ i~\frac{\partial}{\partial {\bf k}} \
u_{\beta {\bf k}}({\bf x}) 
\label{xelem}
\end{equation}
are integrals over the unit cell of the fast components $u_{\alpha {\bf k}} 
({\bf x})$ in the Bloch wave functions $\psi_{\alpha 
{\bf k}} ({\bf x}) = e^{i {\bf k \cdot x}} u_{\alpha {\bf k}} ({\bf x})$. 
In the following, we use the name ${\bf r}_{\alpha \beta}({\bf k})
=\xi_{\alpha \beta}({\bf k})$ for the band off-diagonal elements 
$\alpha\neq \beta$. They are related to the {\it interband} velocity 
elements by 
\begin{eqnarray}
{\bf v}_{\alpha\beta} ({\bf k})=
\langle \alpha, {\bf k}| (-i\hbar/m_e) 
\nabla | \beta, {\bf k}\rangle= \frac{\epsilon_{\beta} 
({\bf k})-\epsilon_{\alpha}({\bf k})} {i\hbar}~
{\bf r}_{\alpha\beta}({\bf k})\ .
\label{vr}
\end{eqnarray}

The electron-phonon matrix elements \cite{Madelung,Mahanb} 
\begin{eqnarray}
M_{\alpha\beta}({\bf k},{\bf k}-{\bf q})& =& M({\bf q})\
\gamma_{\alpha\beta}({\bf k} ,{\bf k}-{\bf q}) \ ,
\nonumber \\
\gamma_{\alpha\beta} ({\bf k},{\bf k}-{\bf q})&=&
\int\limits_{\mbox{\tiny U.C.}} d{\bf x}\
u^*_{\alpha {\bf k}}({\bf x})\ u_{\beta {\bf k}-{\bf q}}({\bf x})
\label{Mkk}
\end{eqnarray}
include the structure factors $\gamma_{\alpha\beta} ({\bf k},{\bf k}-{\bf q})$,
which separately depend on the ``incoming" and ``outgoing" momenta 
$\gamma_{\alpha\beta} ({\bf k}, {\bf k} -{\bf q})\neq f({\bf q})$. Therefore, 
the factor $M({\bf q})$ in the expression (\ref{Mkk}) gives the rate of
electron scattering, while the real space shifts of electrons are 
solely controlled by the lattice \cite{Belinicher,Kurt}, determining 
the factors $\gamma_{\alpha\beta} ({\bf k},{\bf k}-{\bf q})$
(see Appendix B). In the approximation of a constant LO phonon energy 
$\hbar \omega_q \approx \hbar \omega_Q$, the element $M^2({\bf q})$ is given by 
\cite{Mahanb} 
\begin{equation}
M^2({\bf q}) =\frac{M_0^2}{|{\bf q}|^2}\ ,\ \ \ 
M_0^2= 2\pi e^2 \hbar \omega_Q
\left(\frac{1}{\epsilon_{\infty}}-\frac{1}{\epsilon_{0}}\right)\ .
\vspace{2mm}
\label{M}
\end{equation}
The parameters relevant for GaAs are used;  $\hbar \omega_Q=36$ meV, 
$\epsilon_0=12.5$ and $\epsilon_\infty=10.9$.

%%%%%%%%%%%%%%%%%%%%%%%%%%%%%%%%%%%%%%%%%%%%%%%%%%%%%%%%%%%%%%%%%%%%%%%%%%%%%%%
\section{Description of the system}
%%%%%%%%%%%%%%%%%%%%%%%%%%%%%%%%%%%%%%%%%%%%%%%%%%%%%%%%%%%%%%%%%%%%%%%%%%%%%%%
We describe the problem by nonequilibrium Green functions \cite{KaBa,HaJa}
in the matrix form $G_{\alpha \beta}$, similarly as in Ref.\cite{KralCOH}. 
The shift current can be expressed through the {\it interband} elements for 
NGF, which must be known on a semi-analytical level, so that the space 
shifts can be obtained first by algebraic operations. In older works, the 
manipulations have been carried out in the density matrix formalism, but 
the intermediate steps were less clearly presented. More recently, NGF 
were used to investigate the shift current from electron hopping in 
superlattices \cite{Lyanda}, and in other photovoltaic phenomena 
\cite{Sulimov,Spivak}.
 
Here, we develop an approach similar to Ref.\cite{Lyanda}, and 
apply it to our system in a steady-state regime. The interband functions 
$G_{\alpha \beta}$ ($\alpha \neq \beta$) are expressed from the {\it 
differential} version of the Kadanoff-Baym equations \cite{KaBa} (KBE) 
in terms of the intraband $G_{\alpha \alpha}$ from the scattering integrals
of these KBE. Then $G_{\alpha \beta}$ are substituted in the formula for the 
shift current, where the mentioned reordering is performed. Finally, $G_{\alpha 
\alpha}$ are calculated numerically from the {\it integral} version of 
KBE \cite{KralCOH}. We have not been able to perform this reordering,
at least for ${\bf J}_s$, when $G_{\alpha \beta}$ was expressed
from the integral KBE.

%---------------------------------------------------------------------------
\subsection{The differential KBE}
%---------------------------------------------------------------------------
The differential version of KBE for the matrix correlation function 
${\bf G}^<$ is \cite{KaBa,HaJa} ($\chi=({\bf k},t)$, integration over 
$\bar{\chi}$ is implied)
\begin{eqnarray}
\biggr( {\bf G}^{R,0}(\chi,\bar{\chi}) \biggl)^{-1} 
{\bf G}^<(\bar{\chi},\chi^{'})
=\biggl( {\bf \Sigma}(\chi,\bar{\chi})\ {\bf G}(\bar{\chi},\chi^{'}) 
\biggr)^< \ ,
\label{GG}
\end{eqnarray}
where $\bigr( {\bf G}^{R,0}  \bigl)^{-1}= \bigr( G_{\alpha \beta }^{R,0}\ 
\bigl)^{-1} \delta _{\alpha \beta }$ is the inverted free propagator
\cite{KaBa}.  The lesser operation $<$ can be applied to the functions 
on the r.h.s of (\ref{GG}) with the help of the Langreth-Wilkins (LW) rules 
\cite{Langreth} $(A B)^< =A^< B^A+ A^R B^<$. The KBE can be also written 
in a form, where $\bigr( {\bf G}^{A,0}\bigl)^{-1}$ acts on ${\bf G}^<$ 
from the right side, and, in the scattering term, ${\bf \Sigma}$ is 
interchanged with ${\bf G}$. 

Subtraction of these two KBE gives \cite{HaJa} ($T=(t+t^{'})/2$)
\begin{eqnarray}
&&i\hbar\ \frac{\partial G_{\alpha\beta}^<(\chi,\chi^{'})}{\partial T}
-(\epsilon_{\alpha}({\bf k})-\epsilon_{\beta}({\bf k}^{'}))\
G_{\alpha\beta}^<(\chi,\chi^{'})
\nonumber\\
&=&\Sigma_{f;\alpha\bar{\gamma}}(\chi,\bar{\chi})\
   G_{\bar{\gamma}\beta}^<(\bar{\chi},\chi^{'})
  -G_{\alpha\bar{\gamma}}^<(\chi,\bar{\chi})\
   \Sigma_{f;\bar{\gamma}\beta}(\bar{\chi},\chi^{'})\
\nonumber \\
&+& \biggl( \Sigma_{s;\alpha\bar{\gamma}}(\chi,\bar{\chi})\
   G_{\bar{\gamma}\beta}(\bar{\chi},\chi^{'}) \biggr)^<
 - \biggl( G_{\alpha\bar{\gamma}}(\chi,\bar{\chi})\
   \Sigma_{s;\bar{\gamma}\beta}(\bar{\chi},\chi^{'}) \biggr)^<\ .
\label{DIFG}
\end{eqnarray}
In the scattering integrals on the r.h.s., field $\Sigma_{f;\alpha\beta}$ 
and scattering $\Sigma_{s;\alpha\beta}$ self-energy functions are 
introduced \cite{KralCOH}. The terms do {\it not} commute because of 
the time, momentum and band indices. The momentum and band non-commutativity 
ultimately leads 
to the shift currents. The time issue should be less serious, if pulse 
fields with slow envelope functions are applied to the system, where 
gradient expansions in terms of the difference time $\tau$ could be performed 
around the CMS time coordinate $T$ \cite{KaBa,KralGRA}. We adopt this 
approach here and include only the zero order terms. 

%---------------------------------------------------------------------------
\subsection{The self-energy functions}
%---------------------------------------------------------------------------
In order to obtain the field self-energy $\Sigma_{f;\alpha \beta}$
for the Hamiltonian (\ref{h}) in the length gauge, it is necessary to 
perturbatively expand the Green functions in terms of the light excitation,
similarly as in Ref.\cite{KralCOH}. Direct calculation gives 
\begin{eqnarray}
\Sigma_{f;\alpha \beta}({\bf k},{\bf k}^{'},t,t^{'})& =&
- i~e~\delta(t-t^{'})\ \delta({\bf k}-{\bf k}^{'})\
\nonumber \\
& \times&  {\bf E}(t) \cdot
\biggl\{ \frac12\ \biggl( \frac{\partial}{\partial {\bf k}^{'}}
- \frac{\partial}{\partial {\bf k}} \biggr)\ \delta_{\alpha\beta}
- i~\xi_{\alpha\beta}({\bf k}) \biggr\}\ ,
\label{SF}
\end{eqnarray}
which has zero correlation parts $\Sigma_{f;\alpha \beta}^{<,>}
({\bf k},{\bf k}^{'},t,t^{'})=0$. This self-energy acts as an operator, 
due to the presence of derivatives. When surrounded by functions with 
two momenta variables, it differentiates the front (back) function over 
${\bf k}$ (${\bf k}^{'}$), and set ${\bf k}={\bf k}^{'}$. The derivatives 
can be shifted from one side to the other side by a partial integration, 
which changes their signs and combines the two with a prefactor $1$.  
We can also introduce its steady-state form
\begin{eqnarray}
\Sigma_{f;\alpha \beta}^{\pm}({\bf k},{\bf k}^{'})& =&
- i~e\ \delta({\bf k}-{\bf k}^{'})\ {\bf E}_{\pm \omega_0} \cdot
\biggl\{ \frac12\ \biggl( \frac{\partial}{\partial {\bf k}^{'}}
- \frac{\partial}{\partial {\bf k}} \biggr)\ \delta_{\alpha\beta}
- i~\xi_{\alpha\beta}({\bf k}) \biggr\}\ ,
\label{SFSS}
\end{eqnarray}
which can be obtained by a Fourier transform of the Dyson equation in the 
frequency domain \cite{KaBa}. The frequency argument of the Green functions
following $\Sigma_{f;\alpha \beta}^{\pm}$ are shifted by $\mp \omega_0$ 
(see Ref.\cite{KralCOH}).

Finally, we have to write down the correlation function for the 
electron-phonon self-energy, which we use in the self-consistent Born 
approximation
\begin{eqnarray}
\Sigma_{s;\alpha\beta}^<({\bf k},{\bf k}^{'},t,t^{'})& = &
M_{\alpha\gamma}({\bf k}, {\bf k}-\bar{{\bf q}})\
M_{\delta\beta}({\bf k}^{'}-\bar{{\bf q}},{\bf k}^{'})
\nonumber \\
& \times &
G_{\gamma\delta}^<({\bf k}-\bar{{\bf q}},{\bf k}^{'}-\bar{{\bf q}},t,t^{'})\
D^<(\bar{{\bf q}},t,t^{'})\ .
\label{S<}
\end{eqnarray}
Here, the non-locality in momentum and time as well as  all kinds of interband 
transitions are included (summation over $\bar{{\bf q}}$ and repeated band
indices).

%---------------------------------------------------------------------------
\subsection{The total shift current density ${\bf J}$}
%---------------------------------------------------------------------------
The shift current density ${\bf J}$ can be expressed in terms of the {\it 
interband} velocity elements ${\bf v}_{\beta \alpha}({\bf k})$ and the 
${\bf k}$-diagonal elements of the correlation functions 
$G_{\alpha\beta}^<({\bf k},{\bf k})$.  If the perturbing electric 
field is homogeneous in real space, {\it i.e.} invariant with respect 
to the lattice translation of the crystal, then we would naturally expect
that these functions are ${\bf k}$-diagonal $G_{\alpha \beta}^< 
({\bf k},{\bf k}^{'}) = G_{\alpha \beta}^< ({\bf k})\ \delta({\bf k}
-{\bf k}^{'})\ (2\pi)^3$.  In reality, Eqn.(\ref{DIFG}) for the Hamiltonian 
(\ref{h}) in the length gauge should be further transformed \cite{HaJa}, 
to explicitly give zero off-diagonal elements, and this could generate 
additional terms. In our problem these transforms would be complex, so 
that we make the {\it Ansatz} that Eqn.(\ref{DIFG}) gives momentum diagonal
$G_{\alpha \beta}^<$. 

The steady-state shift current density ${\bf J}$ can be expressed as follows
\begin{eqnarray}
\hspace{-10mm}
{\bf J} = 2e  \int \frac{d\hbar\omega}{2\pi}\ 
\int \frac{d{\bf k}}{(2\pi)^3}\
\sum_{\alpha\neq\beta}\ {\bf v}_{\beta \alpha}({\bf k})\
G_{\alpha \beta}^<({\bf k},\omega)
 =  {\bf J_e}+{\bf J_s}+{\bf J_r}\ ,
\label{Jsh}
\end{eqnarray}
where the prefactor 2 encounters the spins. The components ${\bf J_{e,s,r}}$,
which represent contributions of the three processes depicted in 
Fig.\ref{Kral1}a, can be obtained, if ${\bf v}_{\beta \alpha}$
from (\ref{vr}) and the solution $G_{\alpha \beta}^<$ of Eqn.(\ref{DIFG}) 
are substituted in the expression (\ref{Jsh}). In the steady state, the first 
term on the l.h.s. of Eqn.(\ref{DIFG}) is absent. The remaining 
$G_{\alpha \beta}^<$, expressed through the scaterring integrals on the 
r.h.s., can be substituted in (\ref{Jsh}), which was also used in 
Ref.\cite{Lyanda}. In transient situations, the time derivative of  
$G_{\alpha \beta}^<$ in Eqn.(\ref{DIFG}) must be also included, but 
we are postponing this problem to future studies.  Terms with equal band 
indices $\alpha=\beta$ in (\ref{Jsh}) would give the ballistic current 
density, calculated in Ref.\cite{KralCOH} for excitation by two laser beams.  
 
%---------------------------------------------------------------------------
\subsection{The excitation current density ${\bf J_e}$}
%---------------------------------------------------------------------------
Let us first find the expression for ${\bf J_e}$ from the two terms with 
$\Sigma_{f;\alpha\beta}$ on the r.h.s. of Eqn.(\ref{DIFG}). Since ${\bf J_e}$ 
results in the second order of the laser field, $\Sigma_{f;\alpha\beta}$ in 
those terms must be combined with the first order interband correlation 
functions $G_{1;\alpha\beta}^< \approx \Bigl( G_{0;\alpha\alpha}\
\Sigma_{f;\alpha\beta}\ G_{0;\beta\beta} \Bigr)^<$, expressed through 
the interacting Green functions in the absence of laser field  \cite{KralCOH} 
$G_{0;\alpha\alpha}$. The resulting ${\bf J_e}$ has the form
\begin{eqnarray}
{\bf J_e}& = & 2 i e \int \frac{d\omega}{2\pi}\
\int \frac{d\bf{k}}{(2\pi)^3}\
\sum_{\alpha\neq\beta;~\gamma}\ {\bf r}_{\beta\alpha}({\bf k})
\nonumber \\
& \times & \biggl( \Sigma_{f;\alpha\gamma}^{\pm}({\bf k},\bar{\bf k}_1)\
\Bigl( G_{0;\gamma\gamma}(\bar{\bf k}_1,\bar{\bf k}_2,\omega)\
\Sigma_{f;\gamma\beta}^{\mp}(\bar{\bf k}_2,\bar{\bf k}_3)\
G_{0;\beta\beta}(\bar{\bf k}_3,{\bf k},\omega\pm \omega_0) \Bigr)^< \biggr.
\nonumber \\
\biggl.
&-&
\Bigl( G_{0;\alpha\alpha}({\bf k},\bar{\bf k}_1,\omega)\
\Sigma_{f;\alpha\gamma}^{\mp}(\bar{\bf k}_1,\bar{\bf k}_2)\
G_{0;\gamma\gamma}(\bar{\bf k}_2,\bar{\bf k}_3,\omega\pm \omega_0) \Bigr)^<\
\Sigma_{f;\gamma\beta}^{\pm}(\bar{\bf k}_3,{\bf k}) \biggr) \ ,
\label{Je1}
\end{eqnarray}
where we neglect vertex corrections to $G_{1;\alpha\beta}^<$, which 
can contribute especially in transient situations \cite{KralCOH}, and 
assume that $G_{0;\alpha\beta} \approx 0$. We also write two momenta in the 
equilibrium Green functions $G_{0;\alpha\alpha} ({\bf k}, {\bf k}^{'})$, 
in order to perform the derivatives from the field self-energies 
$\Sigma_{f;\alpha\beta} ({\bf k}, {\bf k}^{'})$, even though they 
fulfill $G_{0;\alpha\alpha} ({\bf k}, {\bf k}^{'}) = G_{0;\alpha \alpha} 
({\bf k})\ \delta({\bf k} -{\bf k}^{'})\ (2\pi)^3$ without the above 
Ansatz. In Appendix A, we show \cite{Sipe} that after algebraic manipulations
only elements with $\gamma=\beta$ ($\gamma=\alpha$) remain in the first 
(second) expression of Eqn.(\ref{Je1}). 

In total, we obtain the $c$-component of the vector for the current density 
${\bf J_e}$ 
\begin{eqnarray}
J_e^c& = &2~e^3 
\int \frac{d\omega}{2\pi}\ \int \frac{d\bf{k}}{(2\pi)^3}\ 
\sum_{\alpha\neq\beta}\ 
r_{\beta\alpha;c}^a({\bf k})\ r^b_{\alpha\beta}({\bf k})\
E^a_{\pm \omega_0}\ E^b_{\mp \omega_0} 
\nonumber \\
& \times & \biggl( G_{0;\alpha\alpha}({\bf k},\omega)\ 
G_{0;\beta\beta}({\bf k},\omega\pm \omega_0) \biggl)^< \ ,
\label{Je2}
\end{eqnarray}
where sum over $a$ and $b$ components is performed. The sign convention and 
the terms $r_{\beta\alpha;c}^a({\bf k})$ are described in Appendix A.
For weak scattering, the 
expression (\ref{Je2}) agrees with density matrix calculations \cite{Sipe}.
In this formalism also a nonzero virtual term {\it below} the band gap can
be traced in an expression analogous to (\ref{Je2}). It can be canceled 
by a similar term, with opposite sign, resulting from the band diagonal 
contribution ${\bf v}_{\alpha \alpha}\ G_{\alpha \alpha}^<$ to the current.
% Eq.(37$^{'}$) in Sipe is (\ref{Je2})

For a light linearly polarized in the $b$-direction, the product of the 
components $E^b_{\pm \omega_0}$ with $r_{\beta\alpha;c}^b({\bf k})\ 
r^b_{\alpha\beta}({\bf k})$ can be further simplified. If the 
matrix elements are used in the form \cite{Belinicher,Kurt} 
$r_{\beta\alpha}^b({\bf k})=|r_{\beta\alpha}^b({\bf k})|\ 
e^{i \phi^b_{\beta\alpha}({\bf k})}$, we can arrange ${\bf J}_e$ as follows
\begin{eqnarray}
{\bf J}_e& = &2~e^3 \int \frac{d\omega}{2\pi}\ \int \frac{d\bf{k}}{(2\pi)^3}\
\sum_{\alpha\neq\beta}\ |E^b_{+\omega_0}|^2\
\biggl(i\ {\bf R}^b_{e;\beta\alpha}({\bf k})\ |r^b_{\alpha\beta}({\bf k})|^2
+\frac{1}{2} \frac{\partial~|r^b_{\alpha\beta}({\bf k})|^2}
{\partial {\bf k}}~\biggr)\
\nonumber \\
& \times & \biggl( G_{0;\alpha\alpha}({\bf k},\omega)\
G_{0;\beta\beta}({\bf k},\omega\pm \omega_0) \biggl)^< \ ,
\label{Je22}
\end{eqnarray}
where we have introduced the excitation shift vector ${\bf R}^b_{e;
\beta\alpha}$ with the $c$-component
\begin{equation}
R_{e;\beta\alpha}^{b;c}({\bf k})=
\frac{\partial \phi^b_{\beta\alpha}({\bf k})}{ \partial k^c}
+\xi_{\alpha\alpha}^c({\bf k}) - \xi_{\beta\beta}^c({\bf k})\ .
\label{Rsh}
\end{equation}
From the expression (\ref{Rsh}), it follows that the vector is invariant under 
the phase transformation \cite{Belinicher,Kurt,Sipe} $\psi_{\beta {\bf k}}
({\bf x}) \to e^{i \theta_{n} ({\bf k})} \psi_{\beta {\bf k}}({\bf x})$ of 
the Bloch functions $\psi_{\beta {\bf k}} ({\bf x})$, where $\theta_{n}
({\bf k})$ are arbitrary well-behaved functions. It is also anti-symmetric 
${\bf R}^b_{e;\beta\alpha}({\bf k})= -{\bf R}^b_{e;\alpha\beta}({\bf k})$ 
in the band index $\alpha \leftrightarrow \beta$, since ${\bf r}_{\alpha\beta}
({\bf k})=({\bf r}_{\beta\alpha} ({\bf k}))^{*}$.  Note in (\ref{Je22}), 
that the anti-symmetric ${\bf R}^b_{e;\alpha\beta} ({\bf k})$ combines 
with the imaginary (spectral) values from the Green function product,  
while the band symmetric derivative $\partial |r^b_{\alpha\beta}
({\bf k})|^2 / \partial {\bf k}$ combines with its real (main) part, 
representing renormalization effects due to interactions. An analogous 
situation occurs in the scattering shift vector ${\bf R}_{s;\beta\beta}(
{\bf k}, {\bf k}^{'})$ (see Appendix B) and in the scattering current 
density ${\bf J}_s$, discussed below.

%---------------------------------------------------------------------------
\subsection{The scattering current density ${\bf J_s}$}
%---------------------------------------------------------------------------
The formula for ${\bf J_s}$ can be obtained, if the rest two terms from the 
r.h.s. of Eqn.(\ref{DIFG}), with the electron-phonon self-energy in 
(\ref{S<}), are substituted in the expression (\ref{Jsh}). In this work, 
we consider only {\it intraband} relaxation, so that band-diagonal Green 
functions are used in ${\bf J_s}$. We also assume that they depend on 
one ${\bf k}$ variable, in accordance with our Ansatz.

Let us concentrate first on the two terms ${\bf r}~(\Sigma^R G^< - 
G^R \Sigma^< )$, in the expression resulting from (\ref{Jsh}). They can 
be written as follows
$$
\hspace{-45mm}
\sum_{\alpha\neq\beta}\
{\bf r}_{\beta\alpha}({\bf k})\ \biggl(
 \Sigma_{\alpha\beta}^R({\bf k},\omega)\
 G_{\beta\beta}^<({\bf k},\omega)
-G_{\alpha\alpha}^R({\bf k},\omega)\
 \Sigma_{\alpha\beta}^<({\bf k},\omega)\biggr)
\vspace{-4mm}
$$
$$
= \sum_{\alpha\neq\beta;~\gamma} \hspace{-2mm} {\bf r}_{\beta\alpha}({\bf k})
M_{\alpha\gamma}({\bf k}, {\bf k}-\bar{{\bf q}})
G_{\gamma\gamma}^R({\bf k}-\bar{{\bf q}},\omega-\bar{\omega})
D^>(\bar{{\bf q}},\omega-\bar{\omega})
M_{\gamma\beta}({\bf k}-\bar{{\bf q}},{\bf k})
G_{\beta\beta}^<({\bf k},\omega)
\vspace{-4mm}
$$
$$
- \sum_{\alpha\neq\beta;~\gamma} \hspace{-2mm} {\bf r}_{\beta\alpha}({\bf k})
M_{\alpha\gamma}({\bf k}, {\bf k}-\bar{{\bf q}})
G_{\gamma\gamma}^<({\bf k}-\bar{{\bf q}},\omega-\bar{\omega})
D^R(\bar{{\bf q}},\omega-\bar{\omega})
M_{\gamma\beta}({\bf k}-\bar{{\bf q}},{\bf k})
G_{\beta\beta}^<({\bf k},\omega)
$$
\begin{equation}
- \sum_{\gamma\neq\alpha;~\beta} \hspace{-2mm} {\bf r}_{\alpha\gamma}({\bf k})
G_{\gamma\gamma}^R({\bf k},\omega)
M_{\gamma\beta}({\bf k}, {\bf k}-\bar{{\bf q}})
G_{\beta\beta}^<({\bf k}-\bar{{\bf q}},\omega-\bar{\omega})
D^<(\bar{{\bf q}},\omega-\bar{\omega})
M_{\beta\alpha}({\bf k}-\bar{{\bf q}},{\bf k})\ ,
\label{Js1}
\end{equation}
where the band arguments have been shifted in the last term. In the 
propagator part of the electron-phonon self-energy, the formula 
$\Sigma^R \propto (GD)^R=G^R D^> - G^< D^R$ has been used, which is valid 
for functions with this argument ordering \cite{Langreth,KralGRA}. The 
first term on the r.h.s. of (\ref{Js1}), resulting from  $G^R~D^<$, 
looks similar to the last term, but the two differ in the type 
of phonon correlation functions $D^{>}$, $D^{<}$ and the energy-momentum 
arguments in $G_{\beta\beta}^<$, $G_{\gamma\gamma}^R$ and $M_{\gamma\beta}$.  

This can changed, if the substitution  ${\bf k}^{'}={\bf k} -\bar{{\bf q}}$ 
and $\omega^{'}=\omega-\bar{\omega}$, followed by $\omega^{''}=
-\bar{\omega}$, is used in the last term of (\ref{Js1}); we take advantage 
of the fact that the Bose-Einstein distribution satisfies $n_{BE}(\omega) 
=1/({\rm exp} (\hbar\omega/kT)-1)=-(1+n_{BE}(-\omega))$. The sum of the
first and last terms is
$$
\hspace{-5mm}
\biggl( \sum_{\alpha\neq\beta;~\gamma} {\bf r}_{\beta\alpha}({\bf k})\
M_{\alpha\gamma}({\bf k}, {\bf k}-\bar{{\bf q}})
-\sum_{\gamma\neq\alpha;~\beta} 
M_{\beta\alpha}({\bf k}, {\bf k}-\bar{{\bf q}})\
{\bf r}_{\alpha\gamma}({\bf k}-\bar{{\bf q}})\biggr)\
M_{\gamma\beta}({\bf k}-\bar{{\bf q}},{\bf k})\
$$
\begin{equation}
\times
G_{\gamma\gamma}^R({\bf k}-\bar{{\bf q}},\omega-\bar{\omega})\
D^>(\bar{{\bf q}},\omega-\bar{\omega})\
G_{\beta\beta}^<({\bf k},\omega)\ .
\label{Js2}
\end{equation}
Here the band indices at the electron propagator and correlation function 
can be set equal $\gamma=\beta$, because intraband relaxation is
considered. The commutator in the large bracket, multiplied by 
the element $M_{\gamma\beta} ({\bf k}-\bar{{\bf q}},{\bf k})$, can be 
rewritten in terms of the scattering shift vector ${\bf R}_{s;\beta\beta}
({\bf k},{\bf k}-\bar{{\bf q}})$ and a renormalization correction, as 
shown in Eqn.(\ref{RSD}). If {\it interband} relaxation is also considered,
the resulting formulas can be  generalized analogously.

The other two terms ${\bf r}~(\Sigma^< G^A - G^< \Sigma^A )$ in (\ref{Jsh})
can be reordered similarly, if only the term $G^A D^<$ in $\Sigma^A\propto
(GD)^A=G^A D^> - G^< D^A$ is again considered. 
%Their sum with the expression (\ref{Js2}) is then equal to
%%
%\begin{eqnarray}
%&&2i \sum_{\beta}\ \Biggl\{
%{\bf R}_{s;\beta\beta}({\bf k},{\bf k}-\bar{{\bf q}})\
%|M_{\beta\beta}({\bf k},{\bf k}-\bar{{\bf q}})|^2\ 
%{\rm Im}\ G_{\beta\beta}^R({\bf k}-\bar{{\bf q}},\omega-\bar{\omega})
%\Biggr.
%\nonumber \\
%\Biggl.
%&&- \frac{1}{2} \biggl( \frac{\partial\ 
%|M_{\beta\beta}({\bf k}-\bar{{\bf q}},{\bf k})|^2}{\partial {\bf k}}
%+\frac{\partial\ |M_{\beta\beta}({\bf k}-\bar{{\bf q}},{\bf k})|^2}
%{\partial ({\bf k}-\bar{{\bf q}})}\ \biggr)\
%{\rm Re}\ G_{\beta\beta}^R({\bf k}-\bar{{\bf q}},\omega-\bar{\omega})
%\Biggr\}
%\nonumber \\
%&& \times  D^>(\bar{{\bf q}},\omega-\bar{\omega})\
%G_{\beta\beta}^<({\bf k},\omega)\ .
%\label{Js3}
%\end{eqnarray}
%%
The remaining terms $-~G^<~D^R$, $-~G^<~D^A$ from $(GD)^R$, $(GD)^A$,
respectively, can also be summed to this form. Here, the imaginary part of 
the phonon propagator appears, instead of the electron propagator.
%in Eqn.(\ref{Je22}), (\ref{Js3}). 
Its real part would give further renormalization contributions, 
if phonon scattering was considered.

In total, the scattering shift current density ${\bf J_s}$ 
in steady state is equal to
\begin{eqnarray}
{\bf J}_s & =& -4~e \int \frac{d\omega}{2\pi}\ \int \frac{d\bf{k}}{(2\pi)^3}\
\sum_{\beta}\
\nonumber \\
& \times & \Biggl\{ \Biggl(
{\bf R}_{s;\beta\beta}({\bf k},{\bf k}-\bar{{\bf q}})\
|M_{\beta\beta}({\bf k},{\bf k}-\bar{{\bf q}})|^2\
{\rm Im}\ G_{\beta\beta}^R({\bf k}-\bar{{\bf q}},\omega-\bar{\omega})
\Biggr. \Biggr.
\nonumber \\
& -& \Biggl. \frac{1}{2} \biggl( \frac{\partial\ |M_{\beta\beta}
({\bf k}-\bar{{\bf q}},{\bf k})|^2}{\partial {\bf k}}
+\frac{\partial\ |M_{\beta\beta}({\bf k}-\bar{{\bf q}},{\bf k})|^2}
{\partial ({\bf k}-\bar{{\bf q}})}\ \biggr)\
{\rm Re}\ G_{\beta\beta}^R({\bf k}-\bar{{\bf q}},\omega-\bar{\omega})
\Biggr)
\nonumber \\
& \times& D^>(\bar{{\bf q}},\omega-\bar{\omega})\
G_{\beta\beta}^<({\bf k},\omega)+
{\bf R}_{s;\beta\beta}({\bf k},{\bf k}-\bar{{\bf q}})\
|M_{\beta\beta}({\bf k},{\bf k}-\bar{{\bf q}})|^2\
\nonumber \\
\Biggl. & \times &  \frac{1}{4}\ \Bigl(
 G_{\beta\beta}^<({\bf k}-\bar{{\bf q}},\omega-\bar{\omega})
-G_{\beta\beta}^<({\bf k}-\bar{{\bf q}},\omega+\bar{\omega}) \Bigr)\
 G_{\beta\beta}^<({\bf k},\omega) \Biggr\}\ .
\label{JsT}
\end{eqnarray}
The intraband correlation functions $G_{\beta\beta}^<$ can be obtained from 
the solution of the KBE similarly as in Ref.\cite{KralCOH}. If the 
hot-electron population is small, then it is possible to keep in 
(\ref{JsT}) only the terms linear in $G_{\beta\beta}^<$. 

The renormalization terms in (\ref{JsT}), with derivatives of matrix elements, 
give non-classical corrections to the shift current, due to {\it quasiparticle} 
broadening of the carrier spectra. Note however, that ${\bf J_s}$ is 
practically independent on the scattering rate, at least for weak scattering, 
since ${\bf R}_{s;\beta\beta}$ is determined by the crystal structure 
(see Appendix B), and the relaxation carrier flow is equal to the injected 
flow. From the same reason, the temperature dependence of ${\bf J_s}$ is 
also weak.  Expression analogous to Eqn.(\ref{Je2}), (\ref{JsT}) can be 
derived for the recombination current density ${\bf J_r}$, but in Sec.V 
we argue that ${\bf J_r}$ is usually negligible.

%%%%%%%%%%%%%%%%%%%%%%%%%%%%%%%%%%%%%%%%%%%%%%%%%%%%%%%%%%%%%%%%%%%%%%%%%%%%%%
\section{Approximate solution for the shift currents}
%%%%%%%%%%%%%%%%%%%%%%%%%%%%%%%%%%%%%%%%%%%%%%%%%%%%%%%%%%%%%%%%%%%%%%%%%%%%%%
Here we approximate the expressions for ${\bf J_e}$ and ${\bf J_s}$ to the 
Boltzmann limit, which can give reasonable results in materials with a
moderate scattering. First we simplify ${\bf J_e}$ in (\ref{Je2}), 
where the light orientation is chosen along the (1,1,0) direction, 
appropriate in GaAs. Next, we approximate the equations for the correlation 
functions $G_{\beta\beta}^<$, and insert their solution in Eqn.(\ref{JsT})
for ${\bf J_s}$, together with the shift vector 
${\bf R}_{s;\beta\beta}$ in (\ref{RSD}).

%---------------------------------------------------------------------------
\subsection{The excitation current density ${\bf J_e}$}
%---------------------------------------------------------------------------
In numerical studies, it is convenient to evaluate ${\bf J_e}$ from the
expression (\ref{Je2}) even for a linearly polarized light, so that 
{\it ab-initio} calculations of the elements ${\bf r}_{\beta\alpha}
({\bf k})$ can be performed along the usual crystal axes.  Once evaluated,
${\bf r}_{\beta\alpha}({\bf k})$ can be used to obtain the elements 
${\bf r}_{\beta\alpha;c} ({\bf k})$ from the formulas (3.36-3.37) in 
Ref.\cite{Ghahramani}. At weak scattering, renormalization effects 
contained in the derivatives $\partial |r^b_{\alpha\beta}({\bf k})|^2/
\partial {\bf k}$ from  Eqn.(\ref{Je22}) can be neglected, which, in 
Eqn.(\ref{Je2}), is equivalent to taking only the imaginary part of 
$r_{\beta\alpha;c}^a ({\bf k})\ r^b_{\alpha\beta} ({\bf k})$, anti-symmetric 
in the band indices. Then, $G_{vv}^<~G_{cc}^A$ ($\alpha=v$, $\beta=c$) 
there can be combined with $-G_{cc}^R~G_{vv}^<$ ($\alpha=c$, $\beta=v$), 
which gives $G_{vv}^<~(-2~i~{\rm Im}~G_{cc}^R) =i~A_{vv}~A_{cc}$, 
where $A_{\alpha\alpha}$ are the electron spectral functions \cite{KralCOH}. 
The terms with exchanged band indices are zero, since in equilibrium 
the conduction band is empty; {\it i.e.} $G_{cc}^< ({\bf k},\omega) 
=A_{cc}({\bf k},\omega)\ n_{FD}(\omega) \approx 0$, where $n_{FD}(\omega)
= 1/({\rm exp} (\hbar\omega/kT)+1)$ is the Fermi-Dirac distribution.

The ${\bf k}$ and $\omega$-integration in the two spectral functions
$A_{vv}~A_{cc}$ is performed as in Ref.\cite{KralCOH}.  
A parabolic approximation for the bands is considered in these integrations 
(not in ${\bf r}_{\alpha\beta}({\bf k})$), with wave
vectors tuned to the resonant values ${\bf k}_{res}^n$, used to describe 
electrons that relax between the energy levels $E_{res}^n=E_{res}^0+ 
n\hbar \omega_Q$ (laser tuning to $E_{res}^0$).  The notation 
for angles is related to the electric field ${\bf E}_{\omega_0}=
E_{\omega_0} /\sqrt{2}\ (1,1,0)$; the angle $\phi \in(0,\pi)$ is between 
the direction (1,1,0) and the ${\bf k}_{res}^n$-vectors, and the angle 
$\theta\in(0,2\pi)$ is in the plane orthogonal to the (1,1,0) direction.  
The nonzero $z$-component of the excitation current density is ($\eta= 
(k_{res}^{0},\phi,\theta)$)
\begin{eqnarray}
J_e^z & =&  - i~\frac{ k_{res}^{0} e^3}{\hbar^3}\ \mu_{cv}\ |E_{\omega_0}|^2
\nonumber \\
& \times & 
\int_0^{2\pi} \frac{d\theta}{2\pi}\ \int_0^{\pi} \frac{d\phi}{2\pi}\ 
{\rm sin}(\phi)~ \Bigl( r_{cv;z}^x(\eta)\ r_{vc}^y(\eta)
+r_{cv;z}^y(\eta)\ r_{vc}^x(\eta) \Bigr)\ .
\label{Je3}
\end{eqnarray}
where $\mu_{cv}=m_c m_v/ (m_c + m_v)$ is the effective electron-hole mass,
and the factor in the bracket is purely imaginary. If the light polarization 
is in the (1,-1,0) direction, the signs at both terms with $x$ arguments 
change. Since the integral is the same, the current is opposite, in 
agreement with Fig.\ref{Kral1}b.

%---------------------------------------------------------------------------
\subsection{The scattering current density ${\bf J_s}$}
%---------------------------------------------------------------------------
In order to approximate ${\bf J_s}$ in Eqn.(\ref{JsT}), we need first the
transport equations for the intraband correlation functions $G_{\beta\beta}^< 
({\bf k},\omega)$. Here we describe the hot-electron relaxation by the 
integral KBE derived in Ref.\cite{KralCOH}. In the weak scattering limit, 
they can be approximated to the integral Boltzmann equation (IBE) 
\cite{KralCOH}. For simplicity, we also describe the hot carrier population 
only in the conduction band.  

A laser light polarized along the (1,0,0) direction, used in 
Ref.\cite{KralCOH}, gives a 
nearly isotropic hot-electron distribution in the plane orthogonal to this 
direction, which is sufficiently described by the angle $\phi$. In the 
present work, polarization along the $(1,1,0)$ direction gives a population 
squeezed in the $(0,0,1)$ direction by about $25\ \%$ (see Fig.\ref{Kral2}). 
Moreover, the shift vectors are also very anisotropic (see Fig.\ref{Kral3}b), 
so that both angular variables $\phi$, $\theta$ must be used.  The field
self-energy for {\it interband} transitions along the (1,1,0) direction 
can be obtained from the expression (\ref{SFSS}) in the form
\begin{equation}
\Sigma_{f;cv}^{+}(\phi,\theta)
=-e~\biggl( r_{cv}^x(k_{res}^0,\phi,\theta) 
          + r_{cv}^y(k_{res}^0,\phi,\theta) \biggr)~E_{\omega_0}/\sqrt{2} \ .
\label{SAP}
\end{equation}
In the scattering self-energy $\Sigma_{s;cc}$, the matrix elements 
$M_{cc}({\bf k},{\bf k}^{'})$ use the linearized structure factors 
$\gamma_{cc}({\bf k},{\bf k}^{'})$ from (\ref{ga-kp}). Since {\it intraband} 
relaxation is considered, the square of the exponential prefactor from 
(\ref{ga-kp}) gives $\gamma_{cc}({\bf k},{\bf k}^{'}) \approx 1$, {\it i.e.} 
$M_{cc}({\bf k},{\bf k}^{'})= M({\bf k}-{\bf k}^{'})$ from (\ref{Mkk}). 

Then using the steps in Ref.\cite{KralCOH}, we arrive at the steady-state IBE 
$$
\hspace{-65mm}
f_{cc}(n,\phi,\theta) =
\frac{2\ k_{res}^n\ \tau_o(n)}{\hbar^3}\ \Biggl\{
|\Sigma_{f;cv}^{+}(\phi,\theta)|^2\ \mu_{cv}\ \delta_{n0}
\Biggr.
\vspace{-2mm}
$$
$$
\hspace{17mm}
+\frac{m_c}{2}\ M_0^2\
\int_0^{2\pi} \frac{d\bar{\theta}}{2\pi}\
\int_0^{\pi} \frac{d\bar{\phi}}{2\pi}\ {\rm sin}(\bar{\phi})\
\biggl( \frac{1}{|{\bf k}_{res}^n-{\bf \bar{k}}_{res}^{n-1}|^2}\
f_{cc}(n-1,\bar{\phi},\bar{\theta})\ n_{B}(\omega_Q)
\Bigr.
$$
\begin{equation}
\hspace{-15mm}
\Biggl.  \biggl.  +
\frac{1}{|{\bf k}_{res}^n-{\bf \bar{k}}_{res}^{n+1}|^2}\
f_{cc}(n+1,\bar{\phi},\bar{\theta})\ (1+n_{B}(\omega_Q)) \biggr) \Biggr\}\ ,
\label{IBES}
\end{equation}
where the distribution function on the $n$-th level ($\varepsilon_n^{\pm}
=E_{res}^n \pm n\hbar \omega_Q/2$)
\begin{equation}
f_{cc}(n,\phi,\theta)= \int_{\varepsilon_n^{-}}^{\varepsilon_n^{+}} 
\frac{d\hbar \omega}{2\pi}\ \int_{0}^{\infty} \frac{d k}{2\pi}\ k^2\ 
G_{2;cc}^<(k,\phi,\theta,\omega)
\label{DE2S}
\end{equation}
is defined from the second order $G^< \approx G_{2}^</2!$, expanded in
terms of the interband field self-energy $\Sigma_{f;cv}$; the prefactor 
$2$ in Eqn.(\ref{IBES}) cancels this $2!$. Here $\tau_o(n) =$ 
$-\hbar/(2~{\rm Im} \Sigma_{cc,s}^r(n))$ is the particle relaxation time. 
It is still necessary to add in Eqn.(\ref{IBES}) radiative transfers
of carriers between the bands, but this is practically not reflected in
${\bf J_e}$, and ${\bf J_r}$ is also negligible.

To calculate ${\bf J_s}$, we also need a simpler expression for the 
scattering shift vector ${\bf R}_{s;\beta\beta}({\bf k},{\bf k}^{'})$ in 
(\ref{Rsc}). A tractable approximation \cite{Belinicher,Kurt} can be
obtained if this is linearized in terms of the wave vector difference 
${\bf k}-{\bf k}^{'}$, 
around the center ${\bf k}_0=({\bf k}^{'}+{\bf k})/2$. For large
differences, where the errors can grow, the contributions to scattering
are small, since the elements $|M({\bf k}-{\bf k}^{'})|^2$ decay as 
$|{\bf k}-{\bf k}^{'}|^{-2}$. Direct algebraic manipulation gives the 
scattering shift vector in (\ref{RkO}-\ref{Omega}). In the current 
formula (\ref{JsT}), we expand the vector ${\bf R}_{s;\beta\beta}$ 
in each {\it considered} wave vector ${\bf k}$ in the Brillouin zone. 

Finally, the expression (\ref{JsT}) for ${\bf J}_s$ can be rewritten in the 
Boltzmann limit, where $A_{\beta\beta} ({\bf k},\omega)= -2\ {\rm Im}~ 
G_{\beta\beta}^R({\bf k},\omega) \approx 2\pi \delta (\hbar\omega- 
\epsilon_{\beta}({\bf k}))$.  Since the correlation function $D^>
({\bf q},\omega)$ for free phonons is also sharp, it can be easily 
convoluted over frequency and momentum with $A_{\beta\beta}({\bf k},\omega)$. 
If we consider in (\ref{JsT}) only the term linear in $G_{\beta\beta}^<$,
and neglect renormalization factors, then the approximate ${\bf J}_{s;cc}$ 
results 
\begin{eqnarray}
{\bf J}_{s;cc} & = & -\frac{e~m_{c}\ M_0^2}{\hbar^2}\ \sum_{n}
\int_0^{2\pi} \frac{d\theta}{2\pi}\
\int_0^{\pi} \frac{d\phi}{2\pi}\ \sin(\phi)\
f_{cc}(n,\phi,\theta)\
\nonumber \\
& \times &
\int_0^{2\pi} \frac{d\bar{\theta}}{2\pi}\
\int_0^{\pi} \frac{d\bar{\phi}}{2\pi}\  \sin(\bar{\phi})\ \Biggl(
\frac{k_{res}^{n-1}}{|{\bf k}_{res}^n-{\bf \bar{k}}_{res}^{n-1}|^2}\
({\bf k}_{res}^n-{\bf \bar{k}}_{res}^{n-1})
\times {\bf \Omega}_c({\bf k}_{res}^n)\
(1+n_{B}(\omega_Q)) \biggr.
\nonumber \\
\biggl.
& + &
\frac{k_{res}^{n+1}}{|{\bf k}_{res}^n-{\bf \bar{k}}_{res}^{n+1}|^2}\
({\bf k}_{res}^n-{\bf \bar{k}}_{res}^{n+1})
\times {\bf \Omega}_c({\bf k}_{res}^n)\
n_{B}(\omega_Q) \Biggr)\ .
\label{IBEST}
\end{eqnarray}
Here the distribution from (\ref{DE2S}) has been used, where the integration
over $\omega$ and $k=|{\bf k}|$ from (\ref{JsT}) is already performed.  
A more consistent approximation in Eqn.(\ref{IBEST}) can be obtained 
by evaluating the vectors ${\bf \Omega}_c$ at the points $({\bf k}_{res}^n
+{\bf \bar{k}}_{res}^{\pm 1})/2$. Numerically, it is easier to average 
${\bf \Omega}_c$, evaluated at the side points ${\bf k}_{res}^n$, 
${\bf \bar{k}}_{res}^{\pm 1}$.

%%%%%%%%%%%%%%%%%%%%%%%%%%%%%%%%%%%%%%%%%%%%%%%%%%%%%%%%%%%%%%%%%%%%%%%%%%%%%%%
\section{Numerical results and discussions}
%%%%%%%%%%%%%%%%%%%%%%%%%%%%%%%%%%%%%%%%%%%%%%%%%%%%%%%%%%%%%%%%%%%%%%%%%%%%%%%
In applications, we consider a typical experimental configuration for GaAs; 
the laser intensity is $I=10^6$ W/cm$^2$ at the light energy $\hbar 
\omega_0 =2.1$ eV (band gap $E_g=1.5$ eV), with light polarized along the 
(1,1,0) direction. We calculate the steady-state ${\bf J_e}$ from 
Eqn.(\ref{Je3}), and ${\bf J_s}$ is obtained by solving Eqn.(\ref{IBES}) 
for the intraband distribution $f_{\beta\beta} (n,\phi,\theta)$, which 
is then used in Eqn.(\ref{IBEST}).  The recombination current density 
${\bf J_r}$ is also discussed.  

We consider a model of GaAs with 10 bands, without spin-orbit coupling. 
The matrix elements ${\bf r}_{ij}$ are calculated 
{\it ab initio} within the density functional theory in local density 
approximation using a plane wave-pseudopotential approach \cite{Adolph}. 
They are obtained at the resonant momentum $k_{res}^0$ (resonant value 
for the light energy $\hbar\omega_0$) and at about $40$ energy (momentum) 
levels in the band $c$, corresponding to resonant values of LO 
phonon processes ($E_{res}^n$ at momenta $k_{res}^n$).  The elements are 
found on a mesh $(\phi_i,\theta_j)$, which can be reduced due to symmetry
to a size $10\times 20$ 
points in the region $\phi =(0,\pi/2)$, $\theta=(0,\pi)$, giving in total 
around $100$ Mb of input data.  The calculated current density ${\bf J_e}$ 
corresponds to the excitation between the heavy hole bands ($v=3,4$ in 
Eqn.(\ref{SAP})) and the lowest conduction band ($c=5$), while 
${\bf J_s}$ is illustrated only in the band $c=5$.

%---------------------------------------------------------------------------
\subsection{The electron distributions}
\begin{figure}[htbp]
\vspace*{-55mm}
\hspace*{-48mm}
\epsfxsize=210mm
\epsffile{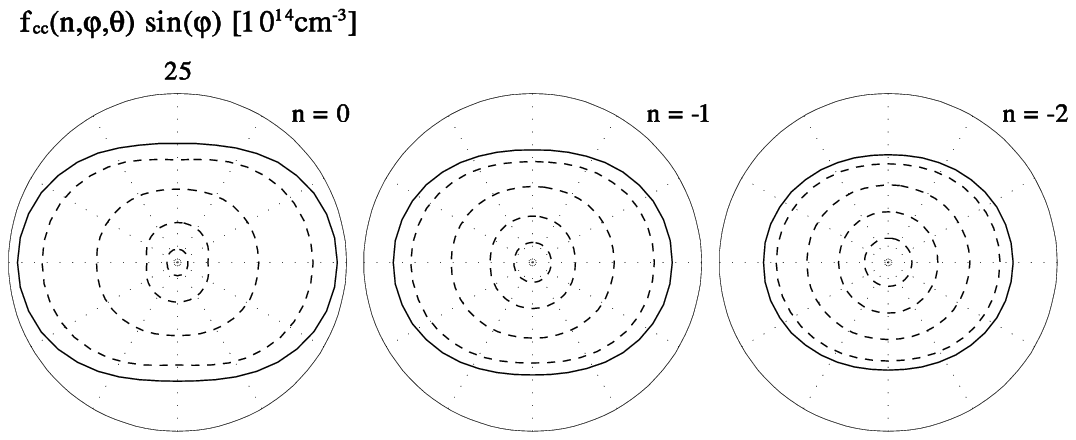}
\vspace{-47mm}
\caption{Crossections through the population $f_{cc}(n,\phi,\theta)$,
multiplied by $\sin(\phi)$,
in the planes orthogonal to the (1,1,0) direction. The values for levels 
$n=0, -1, -2$ show relaxation of the squeezing along the (0,0,1) and 
(1,1,0) directions. The solid and four dashed lines correspond to the 
angles $\phi=i~\pi/10$ ($i=5-1$).}
\label{Kral2}
\end{figure}

%---------------------------------------------------------------------------
In Fig.\ref{Kral2}, we show cross sections, orthogonal to the (1,1,0) direction,
through the electron population $f_{cc}(n, \phi,\theta)$ in the conduction 
band, multiplied by $\sin(\phi)$ as in Eqn.(\ref{IBEST}).  The solid and 
four dashed lines on each plot correspond to the angles $\phi=i~\pi/10$ 
($i=5-1$), while the shapes of the ovals give the $\theta$-distribution. 
The presented levels are $n=0,-1,-2$, and the temperature is $T=300$ K.  
Note, that the population ($n=0$) is squeezed in the (0,0,1) and (1,1,0) 
directions, where the last is seen on the fast decays with $\phi$ of the 
ovals for $n=0$. The population squeezed in the plane orthogonal to the
(1,1,0) direction, around $\phi \approx 0$, contributes more to the shift 
current than the population in the plane orthogonal to the (1,-1,0) direction. 
Therefore, these contributions with opposite signs do not cancel, and
$J_e^z$ can be nonzero. For a light polarized in the (1,-1,0) direction, 
the population is squeezed in the orthogonal direction, according to the 
change in $|r^x_{cv} \pm r^y_{cv}|^2$ from (\ref{SAP}), and $J_e^z$ changes 
sign.  At lower levels both squeezing become relaxed, which visually shrinks 
the size of the population.  Relaxation of anisotropy in hot carrier 
distributions was also studied in Ref.\cite{Binder}.

%---------------------------------------------------------------------------
\subsection{The shift current density ${\bf J_e}$}
%---------------------------------------------------------------------------
In Fig.\ref{Kral3}a, we show the angular dependence of $\Delta_{rr} ={\rm Im}\ 
\bigl(r_{cv;z}^x\ r_{vc}^y $ $+r_{cv;z}^y\ r_{vc}^x \bigr)$, multiplied
by $\sin(\phi)$, from the expression (\ref{Je3}) for ${\bf J_e}$. It is 
calculated for the angles as in Fig.\ref{Kral2}, and for the light 
polarized in the (1,1,0) direction.  If the light is polarized in the 
(1,-1,0) direction, $\Delta_{rr}$ looks the same, but the prefactor 
in Eqn.(\ref{Je3}) changes sign.  This reflects the fact that the 
structure of GaAs allows pumping from different atoms in both situations,
but it gives ${\bf J_e}=0$ for excitation by unpolarized light.  For the 
present excitation, we obtain from Eqn.(\ref{Je3}) the value $J_e^z 
\approx -45$ A/cm$^2$.  This agrees in sign with Fig.\ref{Kral1}b
(negative charge of electrons) and in value with Ref.\cite{Sipe}.

%---------------------------------------------------------------------------
\subsection{The scattering current density ${\bf J_s}$}
%---------------------------------------------------------------------------
Evaluation of ${\bf J_s}$ is sensitive to numerical errors, resulting from 
approximate integrations on the finite mesh for $\phi$, $\theta$.  We have 
corrected three problems in the calculation of ${\bf J}_{s;cc}$.  First, 
it is the flow between level $n$ and $n \pm 1$, second the conservation 
of homogeneous distributions (used as an initial test)
in scattering between level $n$ and $n \pm 1$, and third the fact that 
nonzero shift currents might result even for homogeneous distributions, 
which is a non-physical consequence of the rough mesh and other approximations. 

\begin{figure}[htbp]
\epsfxsize=233mm
\vspace*{-52mm}
\hspace*{-25mm}
\epsffile{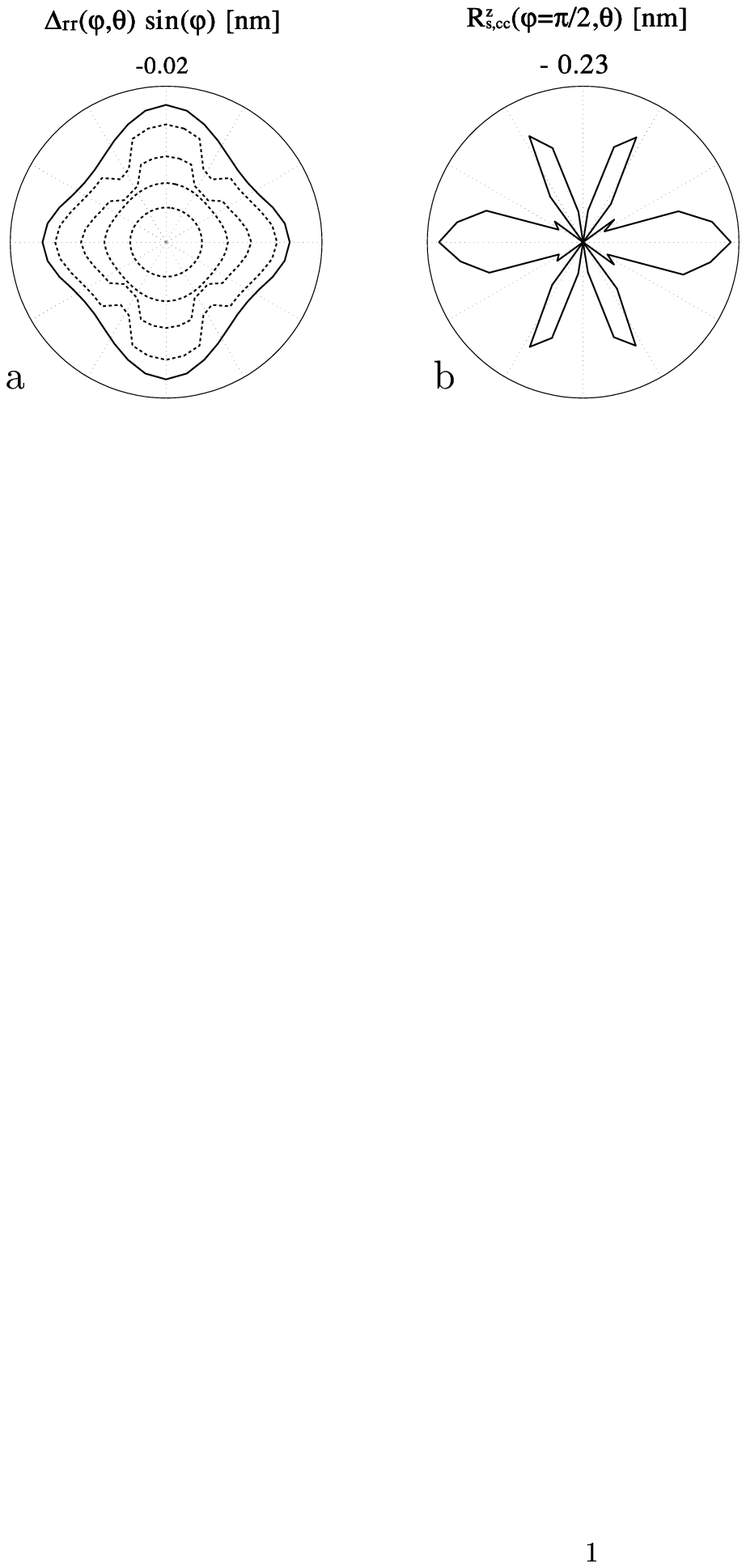}
\vspace*{-220mm}
\caption{Inset a) shows the angular dependence of the factor $\Delta_{rr}
={\rm Im}\ \bigl(r_{cv;z}^x\ r_{vc}^y+r_{cv;z}^y\ r_{vc}^x \bigr)$, multiplied 
by $\sin(\phi)$, calculated from the level $n=0$. The crossections are 
in the plane orthogonal to the polarization direction of light (1,1,0).
It is opposite if light is polarized in the direction 
(1,-1,0).  This proves the mechanism in Fig.\ref{Kral1}b, with zero 
current $J_e^z$ for excitation by unpolarized light.
In the inset b) we present the approximate scattering shift vector 
$R_{s;cc}^z (\phi=\pi/2,\theta) \approx (-{\bf k}\times{\bf \Omega}_c)^z$,
calculated from the level $n=0$. The crossection is as in inset a).
For the orthogonal crossection, the contributions only change sign. 
Thus $J_{s;cc}^z=0$ for homogeneous electron distribution, resulting 
approximately for excitation by unpolarized light. }
\label{Kral3}
\end{figure}

In Fig.\ref{Kral3}b, we show the $z$-component of the linearized scattering 
shift vector $R_{s;cc}^z$ from the level $n=0$; the cross section 
is orthogonal to the (1,1,0) direction at $\phi=\pi/2$.  The behavior of 
$R_{s;cc}^z$ can be appreciated, if we substitute the difference of wave 
vectors in (\ref{RkO}-\ref{Omega}) by $-{\bf k}$. This because, in average, 
the wave vector ${\bf k}$ of the excited electron becomes scattered in the 
$-{\bf k}$ direction, even though the size $k_{res}^0$ of $-{\bf k}$ is 
about an order of magnitude larger than the difference $k_{res}^0- 
k_{res}^{\pm 1}$. The vector $R_{s;cc}^z$ changes a sign,
if the cross section is orthogonal to the (1,-1,0) direction, proving 
zero current for a homogeneous distribution.  Because of the symmetry
specified above, we can limit calculations to one quarter of the Brillouin 
zone. Note that $R_{s;cc}^z$  is larger in the horizontal direction, 
where the squeezed population in Fig.\ref{Kral2} is also larger, so that 
$J_{s;cc}^z$ becomes increased.

\begin{figure}[htbp]
\vspace*{-15mm}
\hspace*{-0mm}
\epsfxsize=165mm
\vspace*{-30mm}
\epsffile{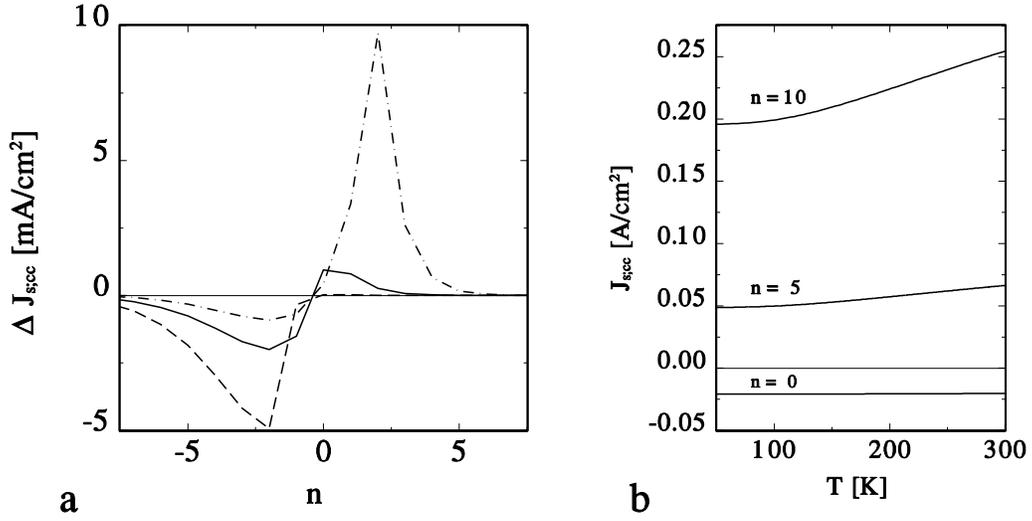}
\vspace*{-10mm}
\caption{Inset a) shows contributions to the scattering shift current
density $J_{s;cc}^z$  from different ``electron levels", due to scattering 
on LO phonons at $T=300$ K. The dashed, solid and dot-dashed lines correspond 
to tuning the laser to the energy of the level $n=-2,0,2$. The change of 
sign of the contributions is due to the fact that $R_{s;cc}^z$ at angles 
$\phi=\pi/2$ changes sign at these excitation energies.  In inset b) we 
present the temperature dependence of the scattering shift current 
density $J_{s;cc}^z$ calculated as a sum of its components in inset a).  
The situations correspond to tuning the laser energy on the levels
$n=0,5,10$. $J_{s;cc}^z$ changes sign and slightly increases with 
temperature, especially at larger $n$.}
\label{Kral4}
\end{figure}

\begin{figure}[htbp]
\vspace*{-25mm}
\hspace*{-0mm}
\epsfxsize=160mm
\epsffile{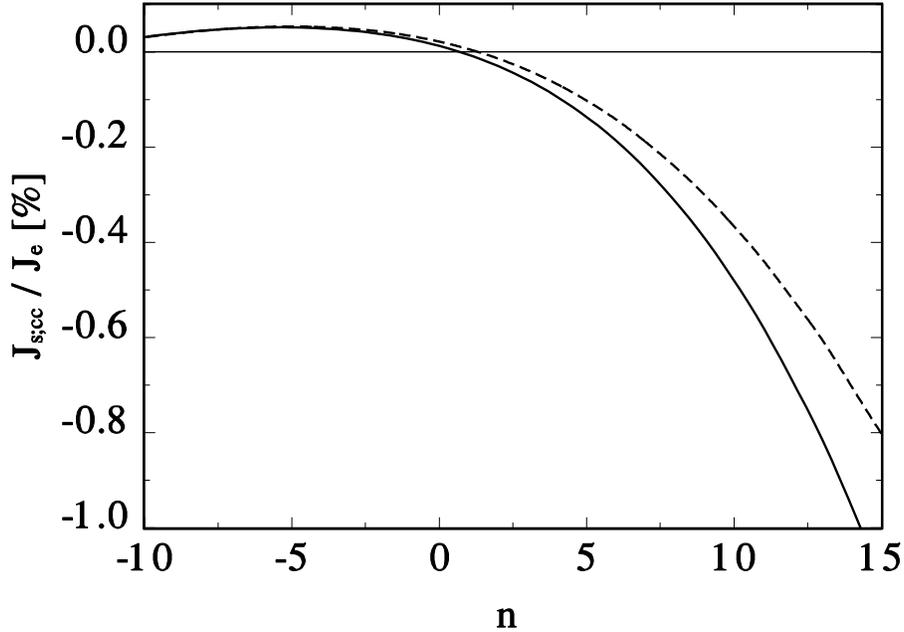}
\vspace*{-18mm}
\caption{Ratio of the scattering and excitation shift current densities
$J_{s;cc}/ J_e$, calculated as a function of excitation energy, labeled 
in phonon levels.  The solid (dashed) lines correspond to $T=300$ K 
($T=50$ K). The ratio changes sign around $n=0$ from positive to negative 
and increases in value up to $1 \%$ for the present excitation energies.}
\label{Kral5}
\end{figure}

In Fig.\ref{Kral4}a, we present contributions to $J_{s;cc}^z$ from different 
electron levels in the conduction band, as induced by phonon scattering 
\cite{KralCOH}. The temperature is $T=300$ K, and the dashed, solid and 
dash-dotted lines correspond to the light excitation at the energies 
corresponding
to the levels $n=-2,0,2$ (used the same injection rate).  The three results 
are very different, since the shift vector $R_{s;cc}^z$ and, consequently,
$J_{s;cc}^z$ changes sign around level $n=0$ (see also Fig.\ref{Kral5}).  
This effect is related to the form of matrix elements, and it appears by 
chance close to the level $n=0$. 

%For excitation at larger $|n|$, the contributions to $J_{s;cc}^z$ have 
%the same sign below and above the excitation point $n$.

The contributions to the current density $J_{s;cc}^z$ are 
summed and presented in Fig.\ref{Kral4}b in the range $T=50-300$ K. The 
excitation energies are labeled by the level number $n=0,5,10$. In this 
interval the response changes sign, as expected from  Fig.\ref{Kral4}a. 
$J_{s;cc}^z$ is about 3 orders of magnitude smaller than $J_e^z$, but 
it increases for larger excitation energies (see Fig.\ref{Kral5}). For 
excitation around $n=0$, $J_{s;cc}^z$ is nearly temperature independent,
while away from $n=0$ it grows in size with $T$. 
In general the increase is small, since $J_{s;cc}^z$ depends weakly on 
the scattering strength (see discussion at Eqn.(\ref{JsT})).

In Fig.\ref{Kral5}, we show the dependence of the ratio $J_{s;cc}^z/J_e^z$ 
on the excitation energy, labeled as  in Fig.\ref{Kral4}b. As already 
mentioned, the ratio is negative close to the $\Gamma$ point, it changes 
sign around $n=0$ and approaches $1\ \%$ at $n=15$. Here we stop our 
calculations, since the departure from the parabolic band approximation 
is large. The solid (dashed) curves correspond to $T=300$ K ($T=50$ K). 
Since $J_e^z$ roughly increases by $25\%$ in this energy region, $J_{s;cc}^z$
is responsible for the increased value in the ratio.  Note also that at 
higher excitation energies the current density $J_{s;cc}^z$ is {\it opposite} 
to $J_e^z$, which could be intuitively expected. The ratio can be larger
for relaxation to/from local side minima in the band structure.

% back to Gamm point > back direction to the original atom.

%---------------------------------------------------------------------------
\subsection{The recombination current density ${\bf J_r}$}
%---------------------------------------------------------------------------
Finally, let us shortly discuss the recombination current density ${\bf J_r}$. 
Since the momentum relaxation time $\tau_p$ is several hundreds fs, and 
the recombination time $\tau_r$ is about 1 ns, only a tinny fraction of 
carriers recombine before randomizing their momenta. The hot carriers 
first isotropically fill the Brillouin zone, and in particular the 
distribution is practically the same in the (1,1,0) and (1,-1,0) directions. 
It is good to recognize that  ${\bf J_e}$ and ${\bf J_r}$ resulting from 
transitions between two particular states have opposite signs. Therefore, 
contributions to ${\bf J_r}$ from the population in the above directions 
are the same in magnitude, but they point in the (0,0,-1) and (0,0,1) 
directions.  This means that ${\bf J_r}$ is practically zero, as suggested 
in Fig.\ref{Kral1}b, where the recombining electrons going from Ga atoms 
do not distinguish between the As atoms. The same observation is expressed 
in Ref.\cite{Sturman} in a more general way, namely, that ${\bf J_r}$ is 
zero in non-pyroelectric materials.

Recombination through trap sites from impurities would give nonzero 
{\it microscopic} shift currents, but its macroscopic value ${\bf J_r}$
averaged over all impurity positions should be negligible.
It would be also interesting to study the shift current related with 
transitions to excitonic levels. Since these levels are energetically
below the band edge, it can be expected that they give different (smaller)
shifts in real space than in the interband transitions. These shifts would
also vary level from level, and ${\bf J_r}$ could be also nonzero.

%%%%%%%%%%%%%%%%%%%%%%%%%%%%%%%%%%%%%%%%%%%%%%%%%%%%%%%%%%%%%%%%%%%%%%%%%%%%%%
\section{Conclusion}
%%%%%%%%%%%%%%%%%%%%%%%%%%%%%%%%%%%%%%%%%%%%%%%%%%%%%%%%%%%%%%%%%%%%%%%%%%%%%%
We have theoretically investigated laser beam generation of the shift current 
density in bulk NCS semiconductors. The excitation, scattering and 
recombination components ${\bf J_{e,s,r}}$ reflect {\it asymmetric carrier 
flows} in elementary cells, induced by the relevant processes. The system 
was described by the NGF methods, which, in combination with
the length gauge, gives a consistent approach to this problem. Expressions 
for ${\bf J_{e,s}}$ were derived for steady-state excitations, in terms 
of the carrier transition rates and the space shift vectors ${\bf R_{e,s}}$. 

For practical purposes, we have simplified the formalism to the Boltzmann 
limit and demonstrated a tractable numerical scheme. Within this 
approximation, we have described optically 
excited GaAs in the presence of LO phonon scattering.  Light induced 
electron transitions between the lowest conduction band and 
the nearest heavy hole bands were considered. For light polarized 
in the (1,1,0) direction, the excitation current density ${\bf J_e}$ in the 
(0,0,-1) direction was obtained. The scattering current density ${\bf 
J_{s;cc}}$ was calculated from the hot-electron distribution in 
the conduction band. It is about two orders of magnitude smaller than 
${\bf J_e}$, and the recombination current density ${\bf J_{r}}$ can 
be neglected, since the momentum relaxation causes that electrons 
recombine with the same strength to atoms placed in opposite directions. 
Therefore, {\it steady-state pumping} of electrons across the crystal
can be realized.

The shift current has potential applications in ultrafast photo-electric
devices, since the response is practically instantaneous.  Modifications 
of the shift current might be observed in low dimensional structures,
heteropolar nanotubes \cite{HETERO} or molecular systems.

\vspace{5mm}
{\bf Acknowledgments}\\
The author would like thank J. E. Sipe for many discussions of the problem 
studied and a financial support provided by Photonic Research Ontario.
He acknowledges B. Adolf and A. Shkrebtii for evaluation of {\it ab-initio} 
matrix elements. The work was greatly improved with the help of K. Busch 
and with comments of A. P. Jauho.

%*************************************************************************
\section*{Appendix A}
\setcounter{equation}{0}
\renewcommand{\theequation}{A.\arabic{equation}}
%*************************************************************************

Let us first compare current expressions from Ref.\cite{Sipe} with our 
results. The sum of the term $\langle J_{intra}^a \rangle^{I}$ and 
$d~\langle {\bf P}_{inter}^{(2)} \rangle/dt$, there, corresponds to 
the excitation shift current density ${\bf J_e}$ from our Eqn.(\ref{Jsh}), 
if scattering is neglected.  The term $\langle J_{intra}^a \rangle^{II}$ 
there corresponds to the excitation part of the ballistic current; 
Eqn.(\ref{Jsh}) with equal bands $\alpha=\beta$, and no transport vertex 
corrections to $G^<_{\alpha\alpha}$. 

Next, we show how Eqn.(\ref{Je2}) can be obtained from Eqn.(\ref{Je1}),
in analogy to Ref.\cite{Sipe}. Two situations can be considered in 
(\ref{Je1}), where either one of the two involved field self-energies 
$\Sigma_{f;\alpha\beta}^{\pm}$ is interband and the other is intraband, 
or both of them are interband. In the first case, the {\it interband}
self-energy in Eqn.(\ref{Je1}) is between the two Green functions, since 
these are from different bands (empty/full), to give real transitions. 
The {\it intraband} self-energy in front or back can be switched by 
partial integration (see comment at (\ref{SF})) to give a derivative, 
with a prefactor $1$, over the front or back variable in the first or 
second term in Eqn.(\ref{Je1}) with $G^<_{1;\alpha\beta}$.  These two can
be combined into a derivative $\partial G^<_{1;\alpha\beta}
({\bf k},{\bf k})/\partial {\bf k}$, and transferred by partial integration 
to the derivative in the prefactor \cite{Sipe} $-\partial {\bf 
r}_{\alpha \beta}({\bf k})/ \partial {\bf k}$. The resulting contribution 
to ${\bf J_e}$ in (\ref{Je1}) is
\begin{eqnarray}
\Delta {\bf J_e}= 2 i e^3 \int \frac{d\omega}{2\pi}\
\int \frac{d\bf{k}}{(2\pi)^3}\ \sum_{\alpha\neq\beta}\ 
\bigl(-i~{\bf r}_{\beta\alpha;a}({\bf k})\ r^b_{\alpha\beta}({\bf k}) \bigr)\
E^a_{\mp \omega_0}\ E^b_{\pm \omega_0}
\nonumber \\
\times \biggl( G_{0;\alpha\alpha}({\bf k},\omega)\
G_{0;\beta\beta}({\bf k},\omega\pm \omega_0) \biggl)^< \  ,
\label{Je1A}
\end{eqnarray}
where we employ the definition \cite{Sipe}
\begin{equation}
{\bf r}_{\beta\alpha;a}({\bf k})=
\frac{\partial {\bf r}_{\beta\alpha}({\bf k})}{ \partial k^a}
-i~\bigl( \xi_{\beta\beta}^a({\bf k}) - \xi_{\alpha\alpha}^a({\bf k}) \bigr)\ 
{\bf r}_{\beta\alpha}({\bf k}) \ .
\label{raba}
\end{equation}
In (\ref{Je1A}) the sum over bands $\alpha\neq\beta$ selects real 
transitions between valence and conduction bands, tuned by the laser 
frequency $\omega_0$. Then, the top (bottom) signs correspond 
to $\alpha=c$, $\beta=v$ ($\alpha=v$, $\beta=c$).

The second situation adds the expression ${\bf r}_{\beta\alpha}({\bf k})\
\sum_{\gamma\neq\alpha,\beta}\ 
\bigl[\xi_{\alpha\gamma}^a({\bf k})\ \xi_{\gamma\beta}^b({\bf k})
-     \xi_{\alpha\gamma}^b({\bf k})\ \xi_{\gamma\beta}^a({\bf k}) \bigr]$
to the bracket $\bigl(-i~{\bf r}_{\beta\alpha;a} 
({\bf k})\ \xi^b_{\alpha\beta}({\bf k}) \bigr)$ in (\ref{Je1A}).
These terms can be largely simplified with the following identity \cite{Sipe}
\begin{equation}
r_{\alpha\beta;b}^a({\bf k})-r_{\alpha\beta;a}^b({\bf k})=
i\sum_{\gamma\neq\alpha,\beta}\  
\bigl[\xi_{\alpha\gamma}^b({\bf k})\ \xi_{\gamma\beta}^a({\bf k})
-     \xi_{\alpha\gamma}^a({\bf k})\ \xi_{\gamma\beta}^b({\bf k}) \bigr]\ ,
\label{ident}
\end{equation}
which can be used to exchange the vector (b) and derivative (a) components 
in $r^b_{\alpha\beta;a}({\bf k})$. If we substitute (\ref{ident}) in
(\ref{Je1A}), and shift arguments, then the square brackets cancel and
the expression (\ref{Je2}) results.

%*************************************************************************
\section*{Appendix B}
\setcounter{equation}{0}
\renewcommand{\theequation}{B.\arabic{equation}}
%*************************************************************************

Here we derive the expression for the intraband scattering shift vector 
${\bf R}_{s;\beta\beta}({\bf k},{\bf k}^{'})$, used in Eqn.(\ref{JsT}),
and linearize it in the wave vector difference ${\bf k}-{\bf k}^{'}$.

We assume \cite{Belinicher,Kurt} that the commutator of the operator 
${\bf x}$ in (\ref{xdefi}) and ${\bf M}$ in (\ref{Mkk}) vanishes, 
{\it i.e}, $ [ {\bf x}, {\bf M} ] = 0$. In the Bloch representation, we 
can easily sum over intermediate states and momenta. Upon separating out the 
terms of the position operator that are band-diagonal, we arrive at 
the following sum rule:

$$
\hspace{-40mm}
\sum_{{\alpha} \neq \beta}\
{\bf r}_{\beta \alpha} ({\bf k}) \,
M_{\alpha \gamma} ({\bf k} , {\bf k}^{'}) -
\sum_{{\alpha} \neq \gamma}\
M_{\beta \alpha} ({\bf k} , {\bf k}^{'}) \,
{\bf r}_{\alpha \gamma} ({\bf k}^{'}) 
$$
\begin{equation}
 = -i \left( \frac{\partial}{\partial {\bf k}} + 
          \frac{\partial}{\partial {\bf k^{'}}} \right) 
M_{\beta \gamma}({\bf k} , {\bf k}^{'})
+ M_{\beta \gamma}({\bf k} , {\bf k}^{'})
 \left(\xi_{\gamma \gamma}({\bf k}^{'}) - \xi_{\beta \beta}({\bf k}) \right)\ .
\label{sumrule}
\end{equation}
%
% Kurt uses x in this commutation (2 k indices + integral over one k), 
% but the off-diagonal part of x with (r \delta) can be integrated over 
% one of the k and gives the above result
%
Aided with  Eqn.(\ref{sumrule}) and writing the matrix elements 
$M_{\beta \beta}({\bf k}, {\bf k}^{'})$ in the form
\begin{equation}
M_{\beta\beta}({\bf k}, {\bf k}^{'})
=|M_{\beta\beta}({\bf k}, {\bf k}^{'})|\ 
e^{i\phi_{\beta\beta}({\bf k},{\bf k}^{'})}\ ,
\label{Mphi} 
\end{equation}
it is now straightforward to simplify the bracketed terms in Eqn.(\ref{Js2})
for $\gamma=\beta$:
$$
\hspace{-15mm}
\biggl( \sum_{\alpha\neq\beta} r_{\beta\alpha}^c({\bf k})\
M_{\alpha\beta}({\bf k}, {\bf k}^{'})
-\sum_{\beta\neq\alpha} 
M_{\beta\alpha}({\bf k}, {\bf k}^{'})\
r_{\alpha\beta}^c({\bf k}^{'})\biggr)\
M_{\beta\beta}({\bf k}^{'},{\bf k})\
$$
\begin{eqnarray}
%&=& M_{\beta\beta}( {\bf k}^{'},{\bf k})\
%\biggl[-i\ \biggl( \frac{\partial}{\partial k^c}
%+\frac{\partial}{\partial k^{'c}} \biggr)
%+\Bigl(\xi_{\beta\beta}^c({\bf k}^{'})-
%\xi_{\beta\beta}^c({\bf k})\Bigr) \biggr]\ 
%M_{\beta\beta}({\bf k}, {\bf k}^{'})
%\nonumber \\
&=& R^c_{s;\beta\beta}({\bf k},{\bf k}^{'})\
|M_{\beta\beta}( {\bf k}^{'},{\bf k})|^2
-\frac{i}{2}\ \biggl( \frac{\partial}{\partial k^c}
+\frac{\partial}{\partial k^{'c}} \biggr)\
|M_{\beta\beta}( {\bf k}^{'},{\bf k})|^2\ .
\label{RSD}
\end{eqnarray}
Here the $c$-component of the scattering shift vector ${\bf R}_{s;\beta\beta}
({\bf k},{\bf k}^{'})$ is 
\begin{eqnarray}
R^c_{s;\beta\beta}({\bf k},{\bf k}^{'}) =\biggl( \frac{\partial}{\partial k^c}
+\frac{\partial}{\partial k^{'c}} \biggr)\ 
\phi_{\beta\beta}({\bf k},{\bf k}^{'})
+ \xi_{\beta\beta}^c({\bf k}^{'})- \xi_{\beta\beta}^c({\bf k}) \ .
\label{Rsc} 
\end{eqnarray}
It has properties similar to the excitation shift vector ${\bf R}_{e;
\alpha\beta}({\bf k})$ in (\ref{Rsh}).  ${\bf R}_{s;\beta\beta} 
({\bf k},{\bf k}^{'})$ is invariant under phase transforms of Bloch 
functions, and anti-symmetric upon exchanging ${\bf k}$ and ${\bf k}^{'}$, 
{\it i.e.} ${\bf R}_{s;\beta\beta} 
({\bf k},{\bf k}^{'})=-{\bf R}_{s;\beta\beta} ({\bf k}^{'},{\bf k})$, which 
follows directly from Eqn.(\ref{Mphi}). Therefore, the real (first) part in 
Eqn.(\ref{RSD}) is also anti-symmetric, while the imaginary (second) part 
is symmetric, and these two give different contributions to the scattering
shift current density ${\bf J}_s$.  

Numerically, it is more convenient to evaluate the shift vector directly 
from from the matrix elements $M_{\alpha\beta}({\bf k}, {\bf k}^{'})$, 
which are accessible to {\it ab-initio} calculations. Then, 
considerable care has to be exerted in order to maintain the invariance 
under phase transformations.  We make use of the specific form of 
$M_{\beta\beta}({\bf k}, {\bf k}^{'})$, with the structure factors 
$\gamma_{\beta \beta}({\bf k},{\bf k}^{'})$ given in Eqn.(\ref{Mkk}), 
to evaluate $R^c_{s;\beta\beta}({\bf k}, {\bf k}^{'})$ from
\begin{eqnarray}
R^c_{s;\beta\beta}({\bf k}, {\bf k}^{'}) 
& = & \mbox{Im} \, \left[
     \frac{1}{{\mid \! \gamma_{\beta \beta}({\bf k},{\bf k}^{'}) \! 
\mid}^{2}}
     \, \gamma_{\beta \beta}^{*}({\bf k},{\bf k}^{'})
     \left(\frac{\partial}{\partial k^c}+
     \frac{\partial}{\partial k^{'c} }
     \right) \gamma_{\beta \beta }({\bf k}, {\bf k}^{'}) \right]
\nonumber \\
    & +& \xi_{\beta \beta}^c({\bf k}^{'}) - \xi_{\beta \beta}^c({\bf k})\ .
\label{RsNEW}
\end{eqnarray}
Since $\gamma_{\beta \beta}({\bf k},{\bf k}^{'})$ only depends on the
band structure, it is clear that the shift vector $R^c_{s;\beta\beta}$
is an {\it intrinsic property of the material} and does not depend on the 
nature of the scattering mechanism. 

Eqn.(\ref{RsNEW}) can be used once $\gamma_{\beta \beta }
({\bf k}, {\bf k}^{'})$ are found.   For practical evaluations it is much 
easier to obtain their linearized form in the difference ${\bf k}-{\bf k}^{'}$.
We can evaluate $u_{\beta {\bf k}}({\bf x})$ and $u_{\beta {\bf k}^{'}}
({\bf x})$ as well as their derivatives from $u_{\beta {\bf k}_0}
({\bf x})$ at ${\bf k}_0=({\bf k}+{\bf k}^{'})/2$ using 
${\bf kp}$-perturbation theory \cite{Pikus}. This allows us to obtain 
corresponding perturbation theoretical
expressions for $\gamma_{\beta \beta }({\bf k}, {\bf k}^{'})$,
$\xi_{\beta \beta}({\bf k})$ and the shift vector 
$R^c_{s;\beta\beta}({\bf k},{\bf k}^{'})$. 

The linearized $u_{\beta {\bf k}}({\bf x})$ results in the form
\begin{equation} 
\hspace{-15mm}
u_{\beta {\bf k}_{0}+\Delta {\bf k}}({\bf x}) =
e^{-i \xi_{\beta \beta}({\bf k}_{0}) \cdot \Delta {\bf k}} \, \left(
u_{\beta {\bf k}_{0}}({\bf x}) - i \, \Delta {\bf k} \cdot
\sum_{\alpha \neq \beta} {\bf r}_{\alpha \beta}({\bf k}_{0}) \,
u_{\alpha {\bf k}_{0}}({\bf x}) \right) \ ,
\label{uDk}
\end{equation}
where $\Delta {\bf k} = {\bf k} -{\bf k}_0$.  The exponential prefactor 
reflects the change of the phase in the Bloch wave $u_{\beta {\bf k}_{0}+\Delta 
{\bf k}}({\bf x})$, which is otherwise completely undetermined.
As mentioned above, this phase is of 
paramount importance, since we have to take derivatives with respect 
to ${\bf k}$ from the Bloch wave $u_{\beta {\bf k}_{0}+\Delta {\bf k}}
({\bf x})$ (see Eqn.(\ref{RsNEW})). Therefore, we have to consider 
{\it all} ${\bf k}$ dependencies of $u_{\beta {\bf k}_{0}+\Delta 
{\bf k}}({\bf x})$ to ensure phase transformation invariance of our results.
In addition, it follows directly from Eqn.(\ref{uDk}) that
\begin{equation}
\frac{\partial u_{\beta {\bf k}} ({\bf x})}{\partial {\bf k}} 
= -i~\xi_{\beta\beta}\ u_{\beta {\bf k}} ({\bf x})
-i \sum_{\alpha\neq\beta} {\bf r}_{\alpha \beta} ({\bf k}) \,  
u_{\alpha {\bf k}} ({\bf x})
\label{uderiv} \ ,
\end{equation}
which is consistent with Eqn.(\ref{xelem}), since the fast component
of Bloch functions $u_{\alpha {\bf k}} ({\bf x})$ and $u_{\beta {\bf k}}^* 
({\bf x})$ are orthogonal in the unit cell.

Using in Eqn.(\ref{RsNEW}) the derivative of the expression 
(\ref{Mkk}), done with the help of Eqn.(\ref{uderiv}), gives the exact result 
\begin{eqnarray}
R^c_{s;\beta\beta}({\bf k}, {\bf k}^{'}) & = &
\frac{1}{\vert \gamma_{\beta \beta}({\bf k}, {\bf k}^{'}) \vert}\
\mbox{Im} \ \left[~i~\gamma_{\beta \beta}^*({\bf k}, {\bf k}^{'})~
{\bf \rho}^c_{s;\beta\beta}({\bf k}, {\bf k}^{'})~\right]\ ,
\nonumber \\
{\bf \rho}^c_{s;\beta\beta}({\bf k}, {\bf k}^{'}) 
& = & \sum_{\alpha \neq \beta} \left( r_{\beta \alpha}^c({\bf k})~
\gamma_{\alpha \beta}({\bf k}, {\bf k}^{'}) -
\gamma_{\beta \alpha}({\bf k}, {\bf k}^{'})~
r_{\alpha \beta}^c({\bf k}^{'} ) \right)\ .
\label{exact}
\end{eqnarray}
Eqn.(\ref{uDk}) may now be used to obtain the {\bf kp}-perturbation 
expression for ${\bf r}_{\beta \alpha}({\bf k})$
\begin{eqnarray}
\hspace{-20mm}
{\bf r}_{\beta \alpha}({\bf k}) & = & e^{i(\xi_{\beta \beta}({\bf k}_0)
-\xi_{\alpha \alpha}({\bf k}_0)) \cdot \Delta {\bf k}} 
\nonumber \\
\hspace{-20mm}
& \times & \left( {\bf r}_{\beta \alpha}({\bf k}_0) + 
\sum_{\gamma \neq \beta} \sum_{\sigma \neq \alpha}
(\Delta {\bf k} \cdot {\bf r}_{\beta \gamma}({\bf k}_0)) \ 
{\bf r}_{\sigma \alpha}({\bf k}_0)\ \delta_{\gamma \sigma} \right) \ .
\label{xi-kp}
\end{eqnarray}
Similarly can be obtained the expression for $\gamma_{\beta \alpha}
({\bf k},{\bf k}^{'})$
\begin{eqnarray}
\gamma_{\beta \alpha}({\bf k},{\bf k}^{'}) & = &
e^{i(\xi_{\beta \beta}  ({\bf k}_0) \cdot \Delta {\bf k}
    -\xi_{\alpha \alpha}({\bf k}_0) \cdot \Delta {\bf k}^{'})}
\nonumber \\
& \times & \left( \delta_{\beta \alpha}
-i~\Delta{\bf k} \cdot \sum_{\sigma \neq \alpha}
{\bf r}_{\sigma \alpha}({\bf k}_0) \ \delta_{\beta \sigma}
+i~\Delta{\bf k}^{'} \cdot \sum_{\beta \neq \sigma}
{\bf r}_{\beta \sigma}({\bf k}_0) \ \delta_{\sigma \alpha} \right)\ ,
\label{ga-kp}
\end{eqnarray}
where $\Delta {\bf k} = {\bf k}  - {\bf k}_0$, $\Delta {\bf k}^{'} = 
{\bf k}^{'}  - {\bf k}_0$.
Upon inserting these results in the expression (\ref{exact}) and making 
use of the vector identity ${\bf a} \times ( {\bf b} \times {\bf c}) 
= {\bf b}~({\bf a}\cdot {\bf c}) - {\bf c}~({\bf a}\cdot {\bf b})$, 
we finally arrive at the linearized expression for the scattering
shift vector
\begin{equation}
{\bf R}_{s;\beta\beta}({\bf k},{\bf k}^{'}) \approx  
({\bf k}-{\bf k}^{'}) \times {\bf \Omega}_{\beta}( {\bf k}_0)\ ,
\label{RkO}
\end{equation}
where the vector ${\bf \Omega}_{\beta}({\bf k}_0)$ is defined as
\begin{equation}
{\bf \Omega}_{\beta}({\bf k}_0)  =  
 \nabla_k \times {\bf \xi}_{\beta\beta}( {\bf k}_0)
= i \sum_{\alpha \neq \beta}\ {\bf r}_{\beta\alpha}( {\bf k}_0) \times
{\bf r}_{\alpha\beta}( {\bf k}_0)\ .
\label{Omega}
\end{equation}
The last expression for $\nabla_k \times {\bf \xi}_{\beta\beta}( {\bf k}_0)$ 
is also derived in Ref.\cite{Aversa} (Eqn.13).  A standard analysis 
using the symmetry properties 
\cite{Blount,Madelung} of the ${\bf r}_{\alpha\beta}({\bf k}_0)$
shows that ${\bf \Omega}_{\beta}({\bf k}_0)$ represents an axial 
pseudo-vector \cite{Kurt} and, as a consequence, can be nonzero only 
if the material does not posses a center of inversion.  It can be 
evaluated numerically from {\it ab-initio} values of the respective 
matrix elements.

\vspace{-10mm}

\end{document}